\documentclass{aa}

\usepackage{amsmath}
\usepackage{txfonts}
\usepackage{aas_macros}
\usepackage{graphicx}

\title{Comparison of two optical cluster finding algorithms \\ for the new generation of deep galaxy surveys}
\author{D. Rizzo\inst{1} \and C. Adami\inst{2} \and S. Bardelli\inst{3} \and A. Cappi\inst{3} \and E. Zucca\inst{3} \and B. Guiderdoni\inst{4} \and G. Chincarini\inst{5,6} \and A. Mazure\inst{2}}

\institute{Universit\`a degli Studi di Milano, via Festa del Perdono 7, 20122 Milano, Italy \and Laboratoire d'Astrophysique de Marseille, 2 place Le Verrier, 13248 Marseille Cedex 4, France \and INAF--Osservatorio Astronomico di Bologna, via Ranzani 1, 40127 Bologna, Italy \and Institut d'Astrophysique de Paris -- CNRS, 98 bis Boulevard Arago, 75014 Paris, France \and Universit\`a degli Studi di Milano-Bicocca, Piazza dell'Ateneo Nuovo 1, 20126 Milano, Italy \and INAF--Osservatorio Astronomico di Brera, via Bianchi 46, 23807 Merate, Italy}

\offprints{Davide Rizzo, \email{rizzo@merate.mi.astro.it}}

\date{Received 2 April 2003 / Accepted 29 September 2003}

\abstract{We present a comparison between two optical cluster finding methods: a matched filter algorithm using galaxy angular coordinates and magnitudes, and a percolation algorithm using also redshift information. We test the algorithms on two mock catalogues. The first mock catalogue is built by adding clusters to a Poissonian background, while the other is derived from N-body simulations. Choosing the physically most sensible parameters for each method, we carry out a detailed comparison and investigate advantages and limits of each algorithm, showing the possible biases on final results. We show that, combining the two methods, we are able to detect a large part of the structures, thus pointing out the need to search for clusters in different ways in order to build complete and unbiased samples of clusters, to be used for statistical and cosmological studies. In addition, our results show the importance of testing cluster finding algorithms on different kinds of mock catalogues to have a complete assessment of their behaviour.
}

\begin{document}

\titlerunning{Comparison of cluster finding algorithms}

\maketitle

\section{Introduction} \label{sec:introduction}

Large and unbiased samples of clusters of galaxies are invaluable tools for investigating cosmology and the large scale structure of the Universe.

Since the compilation of the first optical samples (Abell~\cite{Abell1958}, Zwicky et al.~\cite{Zwicky1961}), it was apparent that selection effects in such catalogues are more difficult to understand and quantify than those in galaxy catalogues. Indeed, although the detection is done on the basis of a galaxy overdensity, the spatial scale and the magnitude of the overdensity vary with the (unknown a priori) redshift. Therefore, other properties of the cluster galaxy population as the  morphology (presence of giant ellipticals) and photometric properties have taken an important role in detections.

Another important problem in detecting optical clusters is the presence of a significant background of field galaxies, which reduces the significance of a detected overdensity, especially at high redshift. A pioneering study on the detection of clusters and its dependence on various selection effects, applying a simple detection algorithm on simulated catalogues, was done by Cappi et al.~(\cite{Cappi1989}); see also van Haarlem et al.~(\cite{vanHaarlem1997}) and Reblinsky \& Bartelmann~(\cite{Reblinsky1999}). Until recently, samples of high redshift clusters were selected almost only in the X-ray band, where this ``background pollution'' is far less important than in the optical band. However, optical cluster samples are still important because the objects are selected on the basis of the stellar light of the galaxy population, thus giving complementary information with respect to the hot gas-X-ray selected clusters. The complementarity of optical and X-ray based searches for clusters has been further reassessed by Donahue et al.~(\cite{Donahue2001}, \cite{Donahue2002}), who showed that these searches sample different cluster populations, only partially overlapping. See also Holden et al.~(\cite{Holden1999}), Adami et al.~(\cite{Adami2000}).

The first automated and objective searches of optical clusters (Dalton~\cite{Dalton1992}, \cite{Dalton1994}, Lumsden et al.~\cite{Lumsden1992}) started when large field galaxies catalogues became available (APM and COSMOS). These searches produced catalogues of nearby clusters.

Recently, more refined statistical techniques, as e.g. the matched filter algorithm (Postman et al.~\cite{Postman1996}) and its refinement EISily (Lobo et al.~\cite{Lobo2000}) were applied to deep imaging surveys like the EIS (Nonino et al.~\cite{Nonino1999}, Scodeggio et al.~\cite{Scodeggio1999}) in order to detect clusters at higher redshift. At the same time, algorithms based on different techniques have been developed, but all based on the detection of some kind of overdensity. It can be an overdensity (or better a sequence) in a colour-magnitude plot, like in the red sequence method (Gladders \& Yee~\cite{Gladders2000}) or an overdensity of photons in the unresolved background, like in the background fluctuations method (Dalcanton~\cite{Dalcanton1996}, Zaritsky et al.~\cite{Zaritsky1997}), which has been used for the Las Campanas Distant Cluster Survey (Gonzalez et al.~\cite{Gonzalez2001}, \cite{Gonzalez2002}). The availability of multiband photometric surveys has encouraged the development of methods making use of colour information, like the already cited red sequence method and the ``cut \& enhance'' method (Goto et al.~\cite{Goto2002}).

Quite surprisingly, little work has been done in order to estimate the relative efficiency and power of different methods in terms of completeness and spurious detections as a function of redshift (see Olsen et al.~\cite{Olsen2001}, Kim et al.~\cite{Kim2002}).

The new generation of redshift surveys, having a high degree of completeness on a wide volume, will permit for the first time the detection of clusters as three dimensional ($\alpha$, $\delta$ and redshift) overdensities, overcoming in part the problem of the high background pollution, and new detection methods have been developed to take advantage of redshift information (e.g. Marinoni et al.~\cite{Marinoni2002}). Indeed, in these cases the main problem is the decrease of the total number of galaxies as a function of redshift, which could be taken into account with the selection function.

A growing number of such surveys is already available or will be started soon, e.g. CNOC2 (Yee et al.~\cite{Yee2000}), SDSS (York et al.~\cite{York2000}), 2dF (Colless et al.~\cite{Colless2001}), VVDS (Le F\`evre et al.~\cite{LeFevre2001}), DEEP II (Davis et al.~\cite{Davis2001}).

In order to avoid biases in subsequent studies, a key information is the selection function of the catalogues produced by the algorithm -- that is, the fraction of detected objects with respect to the total population as a function of richness, redshift and other parameters.

A relatively simple way to find out such information is to create a mock catalogue of galaxies with known characteristics, thus having a complete \emph{a priori} knowledge of the sample of objects we want to investigate. Generally speaking, the simplest way to set up such a catalogue is to build a background of galaxies on which a number of clusters with known richnesses and density profiles are superimposed. Using a simple mock catalogue with known parameters represents a first test for the algorithms, in order to identify the main biases without ambiguity.

On the other hand, the real Universe cannot be simply thought as a superposition of clusters and background galaxies, but includes complex large scale structures such as filaments, ``walls'' and superclusters. Moreover, clusters of galaxies show a huge variety of shapes, profiles and substructures, while a mock catalogue can usually reproduce only a limited range of these parameters. Therefore, more refined tests need more realistic catalogues, as those generated with N-body simulations (see White \& Kochanek~\cite{White2002}, Kochanek et al.~\cite{Kochanek2003} for an application to cluster finding algorithms). Such catalogues come remarkably close to what the real Universe is, as can be seen by computing basic properties such as number counts and angular/spatial correlation function. They also offer a complete knowledge of the galaxy sample, without the additional worries (incompleteness, measuring errors, star/galaxy discrimination) brought about by real surveys. On the other hand, we cannot decide \emph{a priori} the positions and features of the clusters in the sample. The cluster sample can instead be reconstructed \emph{a posteriori}, starting from a quantitative definition of cluster.

The aim of this paper is to investigate the efficiency in detecting clusters and the relative selection effects of the two methods EISily (Lobo et al.~\cite{Lobo2000}) and Spectro (Adami \& Mazure~\cite{Adami2002}). The  EISily algorithm is a purely bidimensional method which uses both overdensity in number of galaxies and a fit to the luminosity function, while the Spectro method works in the combined bidimensional + velocity space.

We first apply the algorithms to a mock catalogue obtained by adding random clusters to a Poissonian background, then on a mock catalogue by Hatton et al.~(\cite{Hatton2003}) generated by N-body simulations. We define three cases with regard to redshift completeness: 100\% completeness down to $I = 24$ (the \emph{complete} sample), 50\% completeness down to $I = 24$ (the \emph{deep} sample), 33\% completeness down to $I = 22.5$ (the \emph{shallow} sample). The last two cases represent reasonable values for the various recent redshift surveys near completion or already available in the literature.

\section{The methods} \label{sec:themethods}

\subsection{The EISily algorithm} \label{sub:eisily}

The EISily algorithm (Lobo et al.~\cite{Lobo2000}) belongs to the \emph{matched filter} category (Postman et al.~\cite{Postman1996}), which has appeared in several ``flavours'' since its introduction. Although originally designed to treat bidimensional data, versions using redshift information have been developed (e.g. Kepner et al.~\cite{Kepner1999}). The characteristics of the code we used for this work is thoroughly described in Lobo et al.~(\cite{Lobo2000}). We will, however, remind its main steps.

EISily is fed with a catalogue of galaxies providing positional data (RA and Dec) and magnitude in one band. Unlike other matched filter algorithms, this one completely separates the \emph{spatial} part (search for significant galaxy overdensities), which comes first, from the \emph{luminosity} part (search of a cluster-like luminosity function). Furthermore, no assumptions are made about the physical size and the exact density profile of clusters. 

For the spatial overdensity detection, a Gaussian filter is applied to a regular grid of points drawn on the catalogue. The grid spacing is equal to the filter angular size $\sigma_\mathrm{ang}$. For each point of the grid, the galaxies within $3\sigma_\mathrm{ang}$ are weighted according to their position, and a S/N ratio is computed by subtracting the signal given by the background (measured in a region of radius $6\sigma_\mathrm{ang}$ around the point) and dividing by the Poissonian standard deviation. A point is flagged as a possible detection whenever its S/N ratio is greater than the S/N of its eight neighbouring points. This test is not performed on points at the edge of the grid, in order to avoid border effects. The S/N ratio of edge points is used only as a comparison value for inner points.

For each candidate cluster detected in this way, a fine centering is then performed. A local grid is applied around each candidate, with a spacing reduced by one half with respect to the original grid. The S/N ratio is computed on the points of this finer grid and compared with the original value. The point with the highest S/N ratio becomes the new provisional cluster position, and the procedure is repeated, with the grid spacing reduced again by one half. Five iterations are made, so that the local grid becomes up to 32 times finer than the original one.

The filter size is then reduced by a factor $\sqrt{2}$. A new, finer grid is applied on the whole catalogue and the steps described above are performed again. Finally, double detections (the ones with separation less than the average of the scales) are removed, leaving only those with greater signal.

We thus end with a preliminary catalogue of candidate clusters, on which a refined background subtraction is made: the background is computed inside a crown around the cluster region. It may happen that the previously detected overdensity disappears, in which case the candidate is removed. No other rejections are made after this step.

Finally the luminosity part of the algorithm is applied. A Schechter function plus power law is fitted to the galaxies of each candidate to simulate the cluster+background luminosity function, and a maximum likelihood analysis (adapted from Schuecker \& B\"ohringer~\cite{Schuecker1998}) is performed, in order to find the best estimate for $m^*$. The slope $\alpha$ remains fixed throughout the computation: we chose a value of $-1.25$.

\subsection{The Spectro algorithm} \label{sub:spectro}

Unlike EISily, this algorithm makes also use of the redshift information. The input parameters are:
\begin{itemize}
\item Minimum and maximum values of Right Ascension, Declination and redshift for the given sample.
\item Angular size of the search window on the sky. This size has to be adapted to the kind of structure we want to detect. For clusters, a good compromise is to adapt this size to match a physical size of R$_{200}$ (typically 1.8 Mpc, $H_0=65$ km~s$^{-1}$~Mpc$^{-1}$). Smaller sizes could be chosen when looking for groups, larger sizes for superclusters or filament searches.
\item Velocity gap for structure analysis, in km~s$^{-1}$ (see e.g. Biviano et al.~\cite{Biviano1997}). This is the optimal velocity gap. For example, a good value for rich cluster searches is 600 km/s. This value is deduced by ranking the galaxies of the sample by increasing redshift, making the histogram of the redshift galaxy-galaxy separation and looking for excesses. In other words, this value is close to the minimum velocity dispersion of the structures to be detected.
\item Minimum number of objects per \emph{elementary} structure. This is the number of objects below which an elementary structure is discarded in the first step of the algorithm (see below). A good value for clusters is five.
\item Minimum number of objects in the \emph{percolated} structure. This is the minimum number of objects in the whole structure (the sum of all the objects detected in each elementary structure). Fixing this number to 10 means that only structures with more than 10 galaxies in total will be kept.
\item Minimum distance between two percolated elementary structures. A good value for clusters is R$_{200}$. If two elementary structures (step 1) are closer than this value, these are assumed to belong to the same ``global'' structure.
\item Maximum extension in redshift of a structure. A good value is six times the typical velocity dispersion of the structures. This parameter must be greater than the gap value.
\end{itemize}

The steps of the algorithm are as follows:
\begin{enumerate}
\item The data are pixelized according to the sky window search and to the velocity gap. If a large enough number of galaxies is found inside a pixel, the coordinates ($\alpha, \delta, z$) of this pixel are stored, as well as the number of galaxies inside this pixel, the minimum and maximum $\alpha, \delta$ and redshift.
\item A percolation algorithm is used to merge elementary structures detected in step 1 in larger isolated structures. The output produces the mean $\alpha, \delta$ and redshift of the structure, an estimate of the number of galaxies inside the structure, its extension in Mpc and in redshift, the minimum and maximum Right Ascension and Declination.
\item A dedicated search of structures is performed around the detected percolated structures. The galaxies inside these structures are listed. For each structure, we have the number of galaxies inside this structure. 
\item A finer analysis of each percolated structure is made, to check for obvious substructures in redshift space. It produces the final list of galaxies associated to this structure.
\end{enumerate}

\section{The simulated catalogues} \label{sec:simulatedcatalogues}

\subsection{The random clusters (RC) mock catalogue} \label{sub:rcmock}

As a first test, the algorithms were applied to an ideal mock catalogue, generated with limits that are typical of the new generation redshift surveys. It is a magnitude limited sample within the range $17 < I \leqslant 24$ covering one square degree. The catalogue was built assuming a $\Lambda$CDM model ($\Omega_M = 0.33$, $\Omega_\Lambda = 0.67$) with $H_0 = 67$ km s$^{-1}$ Mpc$^{-1}$. The space distribution of field galaxies is assumed to be random. Cluster member galaxies were added to clusters with positions randomly distributed in space, following the random cluster (Neymann-Scott) algorithm described by Peebles~(\cite{Peebles1980}). The number of member galaxies is $100 \pm 50$ and uniformly distributed among clusters.  The velocity distribution of member galaxies is Gaussian. Absolute magnitudes were assigned to galaxies following a Schechter luminosity function with $M^* = -22.5$ and $\alpha = -1.25$. After scaling the magnitudes of the member galaxies according to their redshift, those fainter than the magnitude limit of the catalogue were excluded. There were a total of 5\,000 member galaxies, belonging to 523 clusters, which were included in the catalogue: we will refer to these objects as ``sampled members''. Adding these to the 69\,000 field galaxies, we end up with a catalogue made of 74\,000 objects (approximately corresponding to the observed counts). No luminosity evolution was taken into account; a mean K-correction was assigned fitting the values for the $I$ passband given by Poggianti~(\cite{Poggianti1997}).

\begin{figure*}
\centering
\includegraphics[angle=90,width=18cm]{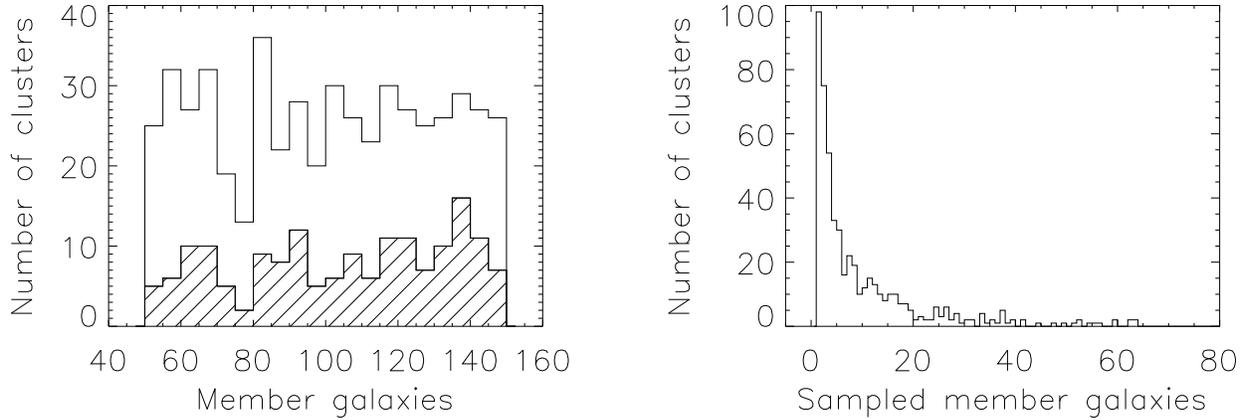}
\caption{Left panel: distribution of cluster richness in the RC catalogue clusters. All the cluster members are taken into account, even those too faint to be included in the final catalogue. The shaded histogram includes those clusters with at least ten members included in the catalogue (\emph{sampled members}). Right panel: distribution of cluster \emph{apparent} richness in the RC catalogue clusters. Only member galaxies above the magnitude limit of the catalogue have been included.}
\label{fig:clusrich}
\end{figure*}

The distribution of cluster richnesses is shown in the left panel of Fig.~\ref{fig:clusrich}. However, these are \emph{all} the member galaxies of the cluster population, including those too faint to be included in the final catalogue. The right panel of Fig.~\ref{fig:clusrich} shows the distribution of \emph{sampled} members for each cluster. It can be seen that several of the 523 clusters of the sample have only one or two sampled member galaxies, which means that they cannot be identified as clusters. Therefore, in the following paragraphs we will also refer to clusters with at least 10 sampled members (166 objects, 32\% of the total sample). The shaded histogram in the left panel shows the clusters with at least 10 sampled members.

\begin{figure}
\resizebox{\hsize}{!}{\includegraphics{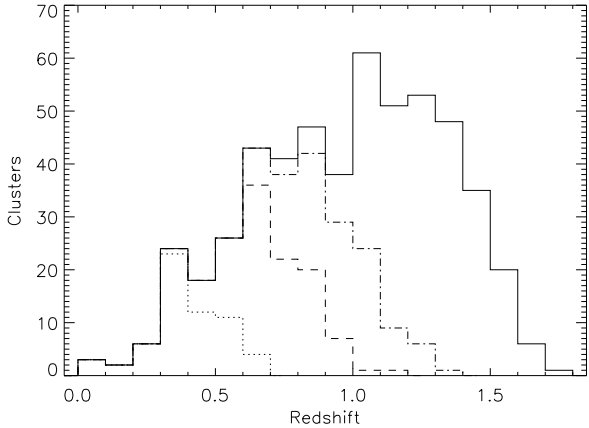}}
\caption{Redshift distribution of input clusters in the RC mock catalogue. Solid line: entire input sample (523 objects). Dashed line: clusters with at least 10 sampled members in the deep sample ($I \leqslant 24$, 166 objects). Dot-dashed line: clusters with at least one member galaxy included in the shallow sample ($I \leqslant 22.5$, 271 objects). Dotted line: clusters with at least ten galaxies included in the shallow sample ($I \leqslant 22.5$, 61 objects).}
\label{fig:simulaoverview}
\end{figure}

Figure~\ref{fig:simulaoverview} show some more data about the catalogue. The solid line histogram is the redshift distribution of all the clusters in the sample. The dashed histogram includes the clusters with at least ten sampled members. The dot-dashed histogram includes clusters with at least one galaxy brighter than magnitude 22.5. Finally, the dotted histogram includes clusters with at least ten galaxies brighter than magnitude 22.5.

\subsection{The N-body mock catalogue} \label{sub:nbodymock}

This synthetic catalogue has been drawn from a first version of the {\sc GalICS} ``hybrid'' model of hierarchical galaxy formation (Hatton et al.~\cite{Hatton2003} and forthcoming papers of the series). The {\sc GalICS} model (for {\it Galaxies in Cosmological Simulations}) uses the outputs of a large, cosmological N-body simulation ($\Lambda$CDM, $256^3$ particles in a box with 150 Mpc on a side) obtained with a parallel TREECODE to follow the merging history trees of dark matter haloes. Then semi-analytic recipes are used to describe gas heating and dissipative cooling, star formation, stellar evolution, stellar feedback to the interstellar/intergalactic medium, and chemical enrichment. The UV/optical/IR/submm spectra of the galaxies are computed from their star formation rate histories, with a simple model of starlight transfer through the dust component, and the {\sc stardust} spectra (Devriendt et al.~\cite{Devriendt1999}). In this scenario, ellipticals and bulges come from galaxy merging. An observing cone is constructed from the {\sc GalICS} outputs by integrating along the line of sight. In this cone, the positions, peculiar velocities, morphological types, magnitudes, colours, and sizes of the galaxies are given by {\sc GalICS}. Here the field covers one square degree. 

The cosmological parameters for the $\Lambda$CDM are $\Omega_M = 0.33$, $\Omega_\Lambda=0.67$, $H_0 = 67$ km s$^{-1}$ Mpc and $\sigma_8h^{-1}=0.88$ (Eke et al.~\cite{Eke1996}). The mass of a single particle is $8.27 \times 10^9 M_\odot$. A friend-of-friend algorithm was run, and haloes were identified where the FOF linked at least 20 particles and the structure was bound ($E<0$). This gives a mass limit for haloes of $1.65 \times 10^{11} M_\odot$, meaning that the big haloes that are galaxy clusters are made from several $10^2$ to $10^3$ such small halos. The mass limit for halos translates into a formal magnitude limit that evolves with redshift, and is $M_B = -18.9$ at $z=0$. The existence of an absolute magnitude limit produces a cutoff in the predicted faint galaxy counts at $R_{AB} = 25.5$. The box of the simulation does not include rare objects, namely those less numerous than 1 per $3.375 \times 10^6$ Mpc$^3$ (modulo the cosmic variance). The most massive halo of the simulation has a mass $M = 7.3 \times 10^{14} M_\odot$.

With a given number of particles (corresponding to given computer resources) the size of the box is a compromise between larger volumes (including rarer objects, but with poorer resolution) and smaller volumes (the rarer objects are missing, but resolution is better). The primary goal of the simulation was to model hierarchical galaxy formation of $L^*$ galaxies with some resolution, hence the choice of the box size. Although there are no haloes as massive as a rich cluster ($\sim 10^{15} M_\odot$), the properties of the galaxies (colours, magnitudes) are in good agreement with data (Hatton et al.~\cite{Hatton2003}).

We used an $I$ selected subsample made of 72\,865 galaxies with $17.0 < I \leqslant 24.0$.

\section{Applying EISily to the catalogues}

We used the same set of initial parameters of Lobo et al.~(\cite{Lobo2000}): an initial $\sigma_\mathrm{ang}$ of 0.024 degrees and five iterations. These values are optimized for finding clusters at $z \la 1.2$.

\subsection{The RC mock catalogue}

Table~\ref{tab:eisilycappi} shows the results of applying EISily to the catalogue. The first two columns show the total number of clusters and the number of clusters with at least 10 sampled members. The three rightmost columns show the number of candidates found by EISily, those with a real counterpart within their estimated angular radius and those without a close counterpart.

\begin{table}
\begin{center}
\begin{tabular}{l l l l l l}
\hline
\hline
Sample & Total & N$_{\geqslant 10}$ & Found & Real & Spurious \\
\hline
All galaxies     & 523 & 166 & 49 & 42 & 7 \\
\hline
\end{tabular}
\end{center}
\caption{Results for EISily on the RC mock catalogue. Column one: number of clusters in the sample. Column two: clusters with at least 10 sampled galaxies. Column three: candidate clusters found by EISily. Column four: candidates with a counterpart closer than their angular radius. Column five: candidates without a close counterpart.}
\label{tab:eisilycappi}
\end{table}

\begin{figure}
\resizebox{\hsize}{!}{\includegraphics{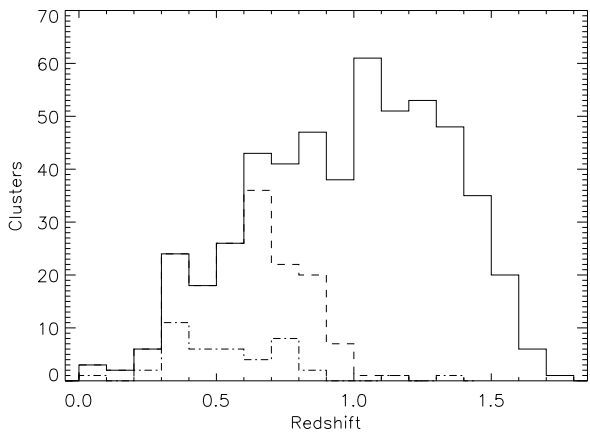}}
\caption{Redshift distribution of clusters found by EISily in the RC mock catalogue. Solid line: entire cluster sample. Dashed line: clusters with at least 10 sampled members. Dot-dashed line: clusters found by EISily with an offset less than their estimated angular radius.}
\label{fig:simulaz}
\end{figure}

Figure~\ref{fig:simulaz} shows the redshift distribution of the clusters found by EISily, compared with the distribution of the entire sample and of the richer clusters, taken from Fig.~\ref{fig:simulaoverview}.

\begin{figure*}
\centering
\includegraphics[angle=90,width=18cm]{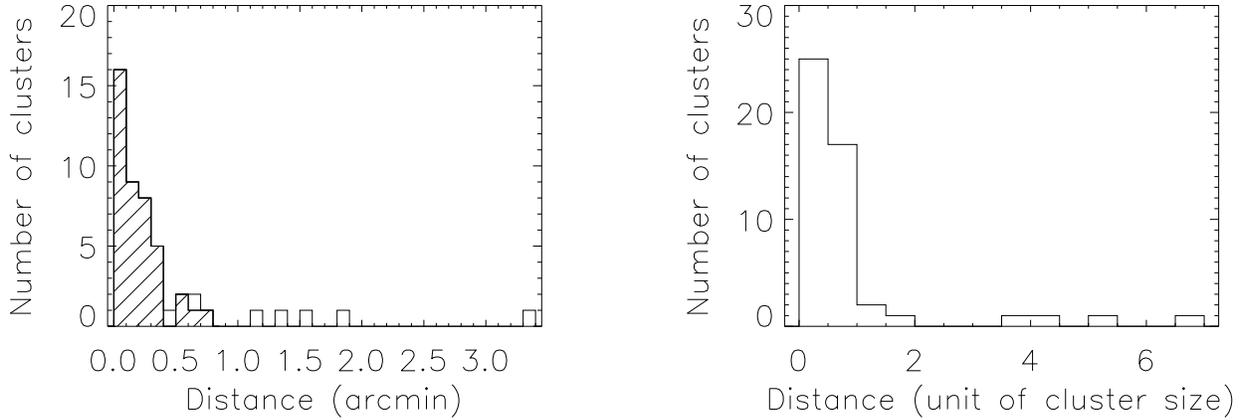}
\caption{These histograms show the distribution of distances between candidates found by EISily and the nearest real cluster. Left panel: distances in arcmin. Right panel: distances in units of the estimated angular radius of the candidate clusters. The dashed histogram in the left panel represents candidates whose distance from the nearest real cluster is less than their angular radius.}
\label{fig:eisilysimuladist}
\end{figure*}

An estimate of the fraction of spurious detections can be done by matching the two catalogues. There are two ways to define the search radius: one is to adopt a constant value for all the clusters, and the other is to use a value proportional to the angular size of each cluster. The algorithm provides an estimate for this parameter (the width of the spatial filter for which the maximum S/N ratio is achieved), so we apply both methods. Results for the entire sample can be seen in Fig.~\ref{fig:eisilysimuladist}. The two histograms show how many candidate clusters have a ``real'' cluster within a certain range of angular distances.

Using a search radius independent of the typical size of the clusters we see that a significant fraction of the sample (25 out of 49 candidates) has a very close ``real'' counterpart (less than 0.2$'$), very unlikely to be the result of random superpositions. We consider this to be a lower limit to the fraction of genuine cluster candidates.

The strong peak observed in the left panel of Fig.~\ref{fig:eisilysimuladist} reappears even stronger in the right panel, where angular distances were computed in units of the angular size of each cluster.

\subsubsection{Redshift and $m^*$ estimates}

EISily offers an indirect method for estimating the redshift of a candidate cluster by comparing the value of $m^*$ given by the code with a typical value of $M^*$ for the photometric band being used. In the case of the RC mock catalogue we know that $M^* = -22.5$ for all the clusters, so we have an opportunity to check the accuracy of the estimate.

\begin{figure*}
\centering
\includegraphics[angle=90,width=18cm]{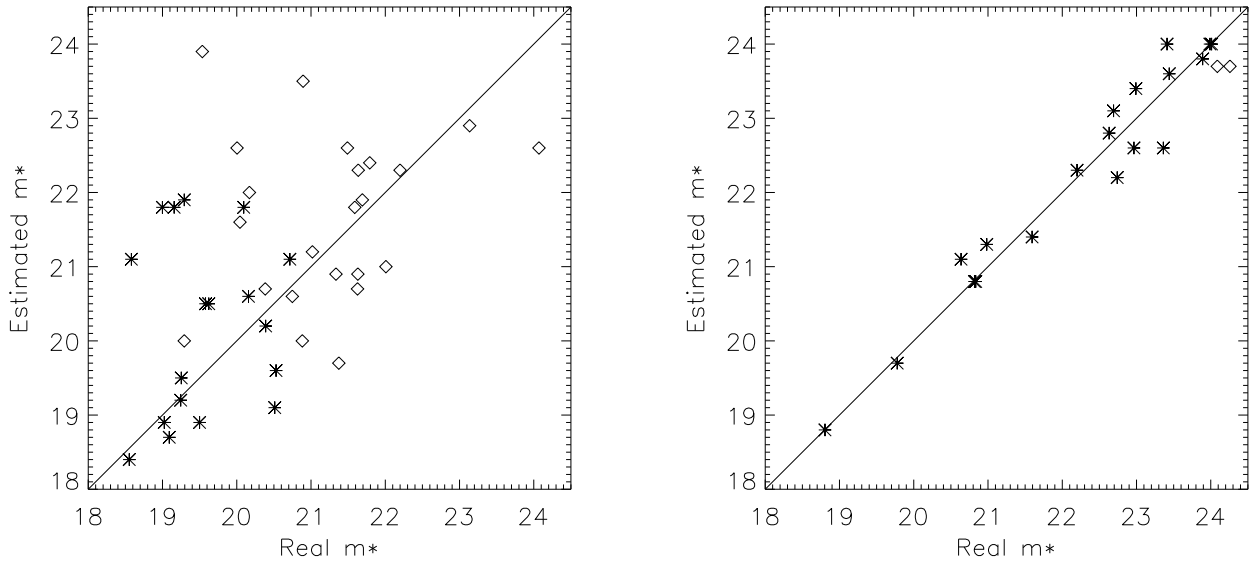}
\caption{Comparison between $m^*$ of the clusters in the RC mock catalogue and $m^*$ fitted by EISily. Stars: clusters with at least 30 sampled members. Diamonds: clusters with less than 30 sampled members. Left panel: original mock catalogue. Right panel: new mock catalogue with richer clusters (30 members brighter than $m^*$).}
\label{fig:mstarcomp}
\end{figure*}

Th left panel of Fig.~\ref{fig:mstarcomp} shows the result of the comparison for the 42 real detected clusters. Stars represent clusters with at least 30 sampled members, while diamonds represent clusters with fewer sampled members. Clusters with more sampled members (which does not necessarily mean intrinsically richer clusters) tend to have brighter $m^*$. This effect has a simple explanation: being $M^*$ fixed for all the clusters, $m^*$ is equivalent to a distance. Distant clusters have dimmer $m^*$ and tend to have fewer member galaxies above the magnitude limit of the catalogue.

As can be seen, the relation shows a great scatter, with a slight tendency of the $m^*$ estimated by the algorithm to be dimmer than the real one. This effect seems not to depend strongly from the number of sampled galaxies of the cluster.

We repeated the test using a catalogue with richer clusters. In this new catalogue, each cluster has 30 members (with a small scatter) brighter than $m^*$, as in the mock clusters used by Lobo et al.~(\cite{Lobo2000}). A cut was applied to the number of observed members, so to have no clusters with more than 400 galaxies. This was to prevent nearby ($z < 0.5$) clusters from showing too many member galaxies (up to 3\,000), which clearly are not observed.

The results of the new test can be seen in the right panel of Fig.~\ref{fig:mstarcomp}. As expected, the scatter in the relation has greatly reduced, and the bias towards fainter magnitudes has disappeared.

As a final remark, we point out that clusters of a given Abell richness have smaller $n^*$ than quoted in Lobo et al., with $n^*$ being the number of galaxies brighter than $m^*$. Lobo et al. assumed that $n^*=30$, 50 and 80  correspond respectively to Abell richness class 1, 2 and 3. But using the same parameters for the luminosity function, we have checked that $n^*=30$ corresponds to about 110 members with magnitudes between $m_3$ and $m_3+2$, i.e. to Abell richness class $R=2$. Analogously, $n^*=50$ corresponds to $R=3$ and $n^*=80$ to $R=4$. Such rich clusters are very rare (and become fewer at higher redshifts), and relatively easy to detect; for this reason we used poorer clusters in our RC catalogue.

\subsection{The N-body mock catalogue}

The code found 20 candidate clusters when applied on the entire sample. This value is to be compared with the 49 candidate clusters found in the RC mock catalogue, which has the same surface and magnitude range, and a similar number of galaxies. It should be noted that the 20 candidates found in the N-body catalogue include both real and spurious ones: without a list of input clusters, there is not a straightforward way to discriminate. That is why this number has to be compared with all the 49 candidates found in the RC mock catalogue, and not just with the 42 real ones.

Since the algorithm was applied with the same initial parameters, this result reflects differences in the two galaxy catalogues. This topic is discussed more deeply in Sect.~\ref{sec:conc}.

\section{Applying Spectro to the catalogues}

\subsection{The RC mock catalogue} \label{sec:spectrorc}

We applied the algorithm to the catalogue, simulating three redshift surveys of different depth and completeness, as explained at the end of Sec.~\ref{sec:introduction}.

\begin{figure*}
\centering
\includegraphics[angle=90,width=18cm]{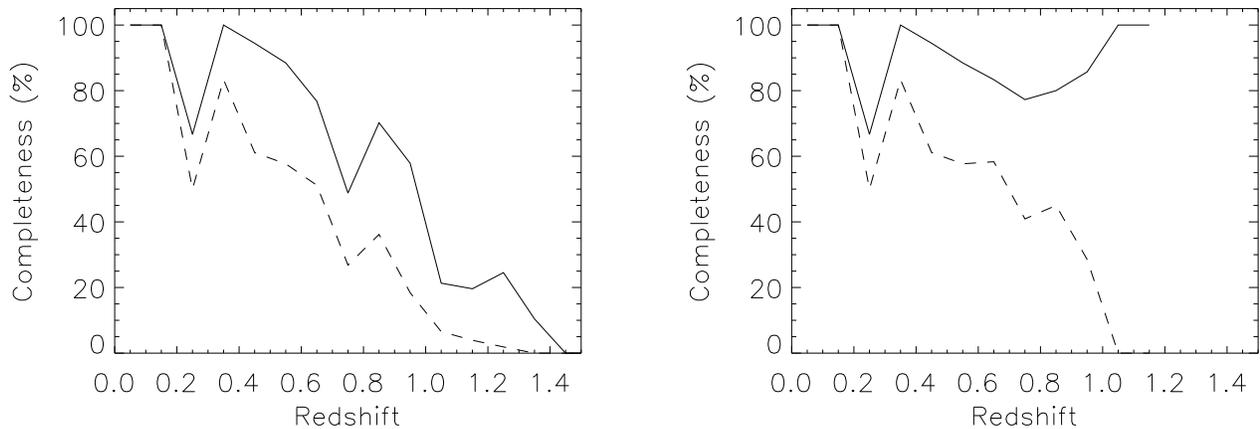}
\caption{Fraction of detected clusters vs. $z$ for the Spectro algorithm applied to the RC mock catalogue. Solid line: complete sample. Dashed line: deep sample. Left panel: fraction of all the clusters of the sample. Right panel: fraction of the clusters with at least 10 sampled members. Spurious candidates are not included.}
\label{fig:adamicappi}
\end{figure*}

\begin{figure*}
\centering
\includegraphics[angle=90,width=18cm]{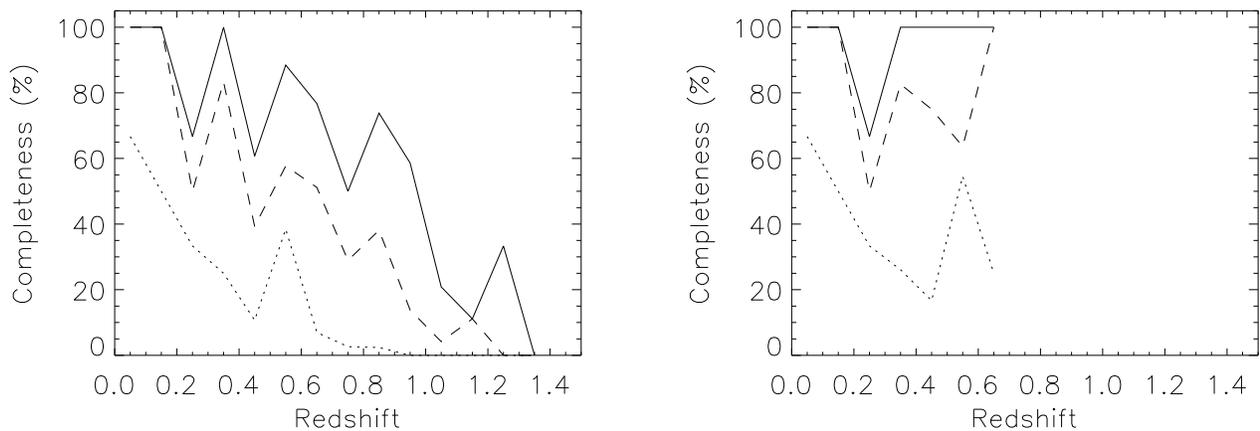}
\caption{Fraction of detected clusters vs. $z$ for the Spectro algorithm applied to the RC mock catalogue. Solid line: complete redshift sampling down to mag 22.5. Dashed line: 50\% sampling down to mag 22.5. Dotted line: shallow sample. Left panel: fraction of the clusters with at least one galaxy brighter than magnitude 22.5. Right panel: fraction of the clusters with at least 10 galaxies brighter than magnitude 22.5. Spurious candidates are not included.}
\label{fig:adamicappibright}
\end{figure*}

In Figs.~\ref{fig:adamicappi} and \ref{fig:adamicappibright} the fraction of detected clusters as a function of redshift is shown for various subsamples. Spurious candidates were \emph{not} included in these figures. The left panel of Fig.~\ref{fig:adamicappi} shows this fraction with respect to all the 523 clusters of the sample. This means that we also included clusters which are undetectable by definition (e.g. those with only one sampled member), which explains why the detection rate progressively worsens for $z > 0.5$. Computing the fraction of detected objects with respect to clusters with at least 10 sampled members, we obtain the results shown in the right panel of Fig.~\ref{fig:adamicappi}. The redshift range is obviously smaller, and it can be seen that the success rate is always above 80\% (within the complete sample), apart from a decrease at $z = 0.2 - 0.3$ due to small statistics. It should also be noted that \emph{all} the clusters up to $z = 0.6$ have more than 10 sampled members (see Fig.~\ref{fig:simulaoverview}), which is why the curves in the two panels of Fig.~\ref{fig:adamicappi} are identical up to that point. 

Figure~\ref{fig:adamicappibright} performs the same analysis for the shallow sample ($I \leqslant 22.5$). First clusters with at least one galaxies brighter than $I = 22.5$ are considered, then those with at least 10 galaxies included in the shallow sample. It can be seen from Fig.~\ref{fig:simulaoverview} that the two subsamples are identical up to $z = 0.4$.

With regard to spurious detections, there were 30 in the complete sample, 17 in the deep sample and 4 in the shallow sample. Expressing these number as fractions of the total number of detections (genuine + spurious) we have 11.9\%, 12.6\% and 11.4\%, respectively. In other words, the false positive rate does not depend on the redshift completeness, at least in the range we explored.

\subsection{The N-body mock catalogue}

\begin{figure}
\resizebox{\hsize}{!}{\includegraphics{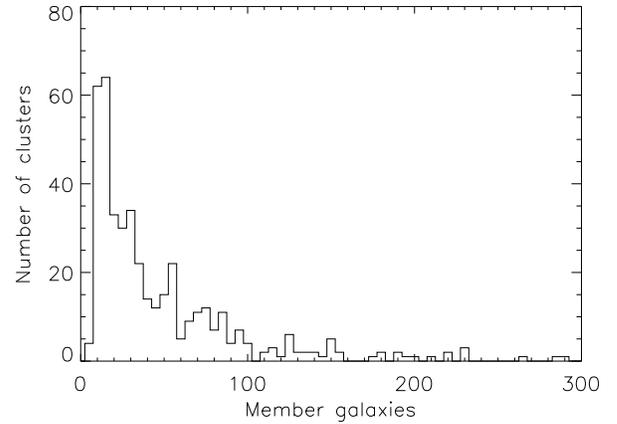}}
\caption{Distribution of the number of member galaxies for the candidate clusters found by Spectro in the N-body mock catalogue.}
\label{fig:structmembers}
\end{figure}

The algorithm found 426 structures: Fig.~\ref{fig:structmembers} shows the distribution of the number of member galaxies. 143 of them (36\%) have less than 20 members.

\begin{figure*}
\centering
\includegraphics{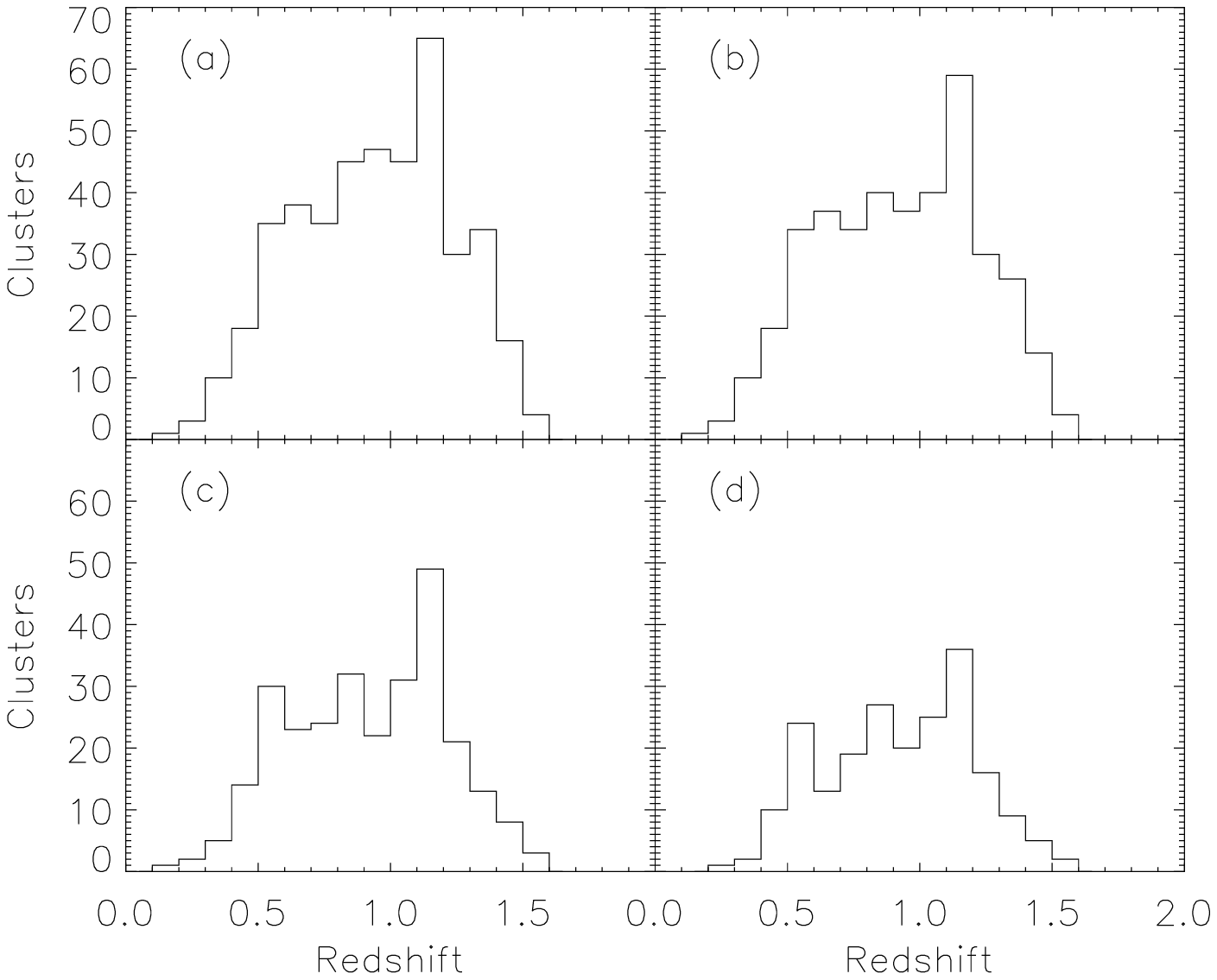}
\caption{Redshift histograms of the clusters found by Spectro. (a) Whole sample. (b) Structures with more than 10 members. (c) Structures with more than 20 members. (d) Structures with more than 30 members.}
\label{fig:speczhist}
\end{figure*}

Figure~\ref{fig:speczhist} shows the redshift histograms of four samples of structures detected by Spectro. Panel (a) was produced using the whole sample. The number of structures per bin increases more or less steadily till the absolute maximum at $1.1 < z < 1.2$, after which there is a decline. This is the same redshift distribution as the real clusters in the RC mock catalogue. Panels (b), (c) and (d) list only the structures with more than 10, 20 and 30 member galaxies, respectively: the histogram is qualitatively the same as before, with the exception of two secondary peaks at $0.5 < z < 0.6$ and $0.8 < z < 0.9$ which became prominent in panels (c) and (d).

\section{Comparing the algorithms}

\subsection{The RC mock catalogue}

We begin by summarizing the results of the two algorithms:
\begin{itemize}
\item EISily found 49 candidates;
\item Spectro found 252 candidates within the complete sample, 135 candidates within the deep sample, 35 candidates within the shallow sample.
\end{itemize}

We cross correlated the catalogues of candidate clusters given by the two algorithms. Table~\ref{tab:cappimatches} lists the results. EISily found 49 candidates (\emph{total} sample), 42 of which had a real cluster closer than their estimated angular radius (sample of \emph{real} candidates). Obviously this result does not depend on redshift sampling. Within the total sample Spectro found 252 candidates, 222 of which had a close real counterpart. The cross-correlation between the 49 candidates by EISily and the 252 candidates by Spectro gave 40 matches (i.e. with distance less than the estimated angular radius of EISily clusters). The cross-correlation between the 42 ``real'' candidates by EISily and the 222 by Spectro gave again the same 40 matches. This means that no matches originated from the random superposition of spurious candidates. 

Within the deep sample Spectro found 135 candidates, and the cross-correlation with the 49 EISily candidates gave 28 matches. 118 candidates by Spectro had a close real counterpart; in the cross-correlation with the 42 ``real'' EISily candidates all the 28 matches appeared again.

The calculation was repeated for the shallow sample. The results are again shown in Table~\ref{tab:cappimatches}.

\begin{table}[htbp]
\begin{center}
\begin{tabular}{l l l l}
\hline
\hline
Sample & EISily & Spectro & Combined \\
\hline
100\%, total       & 49 & 252 & 40 \\
100\%, no spurious & 42 & 222 & 40 \\
50\%, total        & 49 & 135 & 28 \\
50\%, no spurious  & 42 & 118 & 28 \\
33\%, total        & 49 &  35 & 14 \\
33\%, no spurious  & 42 &  31 & 14 \\
\hline
\end{tabular}
\end{center}
\caption{Cross-correlation between the catalogues of candidate clusters generated by EISily and Spectro. Values are reported for three different redshift samplings (100\%, 50\%, 33\%, see Sec.~\ref{sec:spectrorc}). For each sampling, we consider both \emph{all} the candidates found by the algorithms (``total'') and those with a real counterpart closer than their estimated angular radius (``no spurious'').}
\label{tab:cappimatches}
\end{table}

\subsection{The N-body mock catalogue}

The results of the two algorithms applied to the N-body catalogue show a much greater difference than with the RC mock catalogue:
\begin{itemize}
\item EISily found 20 candidates;
\item Spectro found 426 candidates.
\end{itemize}

The cross correlation between the two catalogues, with a search radius of one arcminute, resulted in 8 matches.

For the N-body catalogue we do not have a list of ``real'' input clusters. Therefore, there is not a straightforward way to decide whether a candidate cluster is real or not, as there was with the RC mock catalogue. As a consequence, we can not immediately say which of the 8 matches we found are the result of chance superpositions (between two spurious clusters or a real and a spurious cluster). In order to provide a statistical answer for this question, we repeated the cross-correlation using random lists of candidate clusters. Each catalogue of candidate clusters was randomized 100 times, thus creating 100 random catalogues for each algorithm. The search for matches was repeated for all the couples of random catalogues, finding on average $6.4 \pm 2.0$ matches.

The cross-correlation was redone by taking only the richer objects found by Spectro, namely the ones with more than 20 members (278 objects). The correlation between the original catalogues of candidate clusters resulted in 6 matches, while the random catalogues gave $4.2 \pm 1.8$ matches.

In both cases, matches from the original catalogues of candidate clusters are one sigma higher than the matches from random catalogues. We must therefore conclude that most matches are the result of chance superpositions.

\section{Conclusions} \label{sec:conc}

We have analyzed two cluster finding algorithms with extensive tests on two simulated catalogues. To keep the analysis as unbiased as possible, we did not try to tune the parameters of each method to maximize the correspondence between the results, but we preferred to use two sets of parameters which could be inferred from the physics of cluster populations.

The two mock catalogues had the same surface and magnitude range, and a similar number of objects (74\,000 for the RC mock catalogue and $\sim$ 73\,000 for the N-body mock catalogue).

\subsection{The Spectro method}

The Spectro method recovers a large part of the theoretically detectable clusters up to z$\sim$0.7. For example, within the complete sample we should detect almost all the clusters with more than 10 member galaxies brighter than mag 24. The right panel of Fig.~\ref{fig:adamicappi} shows that the percentage of missed rich clusters is lower than 20\% for any redshift. Even considering poor clusters (with less than 10 members brighter than mag 24), we still detect more than 80\% of the clusters up to z$\sim$0.6 (Fig.~\ref{fig:adamicappi}, left panel). 

Within the deep sample (whose redshift completeness will be typical of new generation deep redshift surveys), the detection method is still efficient: Fig.~\ref{fig:adamicappi} (right panel) shows that the percentage of missed rich  clusters is lower than $\sim$40\% up to z$\sim$0.65.

Finally, assuming shallower redshift surveys (magnitude limit of 22.5), the detection method is still efficient for theoretically detectable clusters (those with at least one galaxy brighter than 22.5). The left panel of Fig.~\ref{fig:adamicappibright} shows that the percentage of missed clusters is lower than $\sim$40\% up to $z \sim 0.7$ (for 100\% redshift sampling).

\subsection{Comparison with EISily for the simulated catalogues}

EISily found 49 objects in the RC mock catalogue and 20 in the N-body mock catalogue, while Spectro found 252 objects in the first case and 426 in the second (with 100\% spectral coverage). This opposite behaviour is probably a consequence of the way the mock catalogues were created. The RC mock catalogue is made by a Poissonian background plus superimposed clusters. It does not reproduce physical effects like the correlation function of galaxies, and lacks structures like filaments and walls, which are indeed overdensities with respect to the field, but are not considered ``clusters'' in a strict sense. Therefore, a 2-D method like EISily is more efficient in the RC mock catalogue, because the assumptions on which it is based (detecting overdensities on a random background) are the same on which the catalogue has been built. In addition, it must be remembered that the N-body catalogue lacks very massive haloes (see Sec.~\ref{sub:nbodymock}), on which EISily is most efficient.

\begin{figure}
\resizebox{\hsize}{!}{\includegraphics{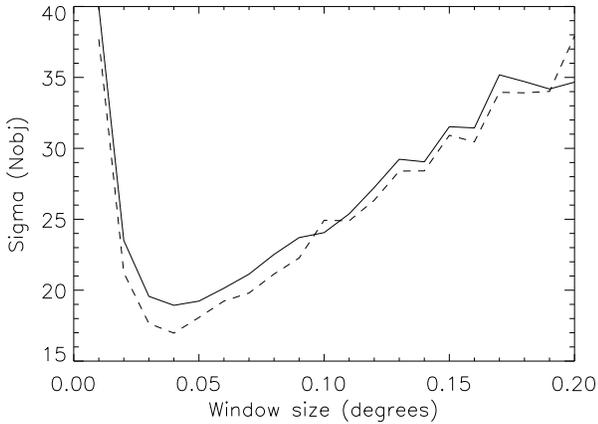}}
\caption{Standard deviation of the number of objects inside a window of a given size, moved on a catalogue. Solid line: RC mock catalogue. Dashed line: N-body mock catalogue. The window has been moved by steps of half its size until the whole catalogue has been covered.}
\label{fig:winstddev}
\end{figure}

It is also relevant the fact that a N-body mock catalogue, especially if deep, has the effect of smoothing and diluting overdensities. This can be simply seen by counting the number of objects inside a window located in different positions of the catalogue, and taking the standard deviation of the result, normalized to the average number of objects inside the window. The result is shown in Fig.~\ref{fig:winstddev}. The N-body mock catalogue is smoother at nearly all scales: the effect is relatively small but systematic.

With regard to Spectro, by making use of the third dimension it can locate clumps and substructures in filaments. Such complex structures are by definition absent in the RC mock catalogue, which can explain the higher number of candidate clusters found in the N-body mock catalogue. It should also be noted that clusters are located at the nodes of filaments and sheets. As a consequence, we expect the EISily method to be less efficient with the N-body catalogue because the surroundings of the clusters have a density that is higher than the true background.

These results suggest the importance of testing cluster finding algorithms on different kinds of mock catalogues to have a complete assessment of their behaviour.

\subsection{Strategy}

The fraction of structures detected by EISily which are also detected by spectro in the RC mock catalogue depends on the redshift completeness. Table~\ref{tab:cappimatches} shows that this fraction goes from 33\% (for the shallow sample) to 95\% (for the total sample). For the N-body mock catalogue the percentage is at most 40\%, assuming that all the matches are not chance ones (which is unlikely); however, let us remind that a 2-D method like EISily is not very efficient on such catalogues.

Our results constitute a case for the complementarity of EISily and Spectro. On the one hand, EISily is more targeted at rich and moderately rich clusters, and can be applied to a catalogue with a single photometric band and no spectroscopic data. For that reason it is useful in the preliminary stage of a spectroscopic survey, as soon as photometric data become available, to select targets for spectroscopic observations that are very likely to be present in the final catalogue created by Spectro. On the other hand, Spectro needs a certain degree of redshift information, which can be achieved by spectroscopy or multiband photometry (with photometric redshifts). While becoming applicable in a slightly later stage of the survey with respect to EISily, it can detect smaller physical structures, some of which at large distances, which would be washed out in a two dimensional view, and can also deproject several structures at different redshifts on the line of sight. This puts in evidence the necessity to use various detection algorithms to avoid to miss entire cluster or structure populations.

With the combination of these -- and other -- algorithms for cluster detection it is possible to produce a large and reliable sample of cluster candidates. The exact nature of each object can be later determined by means of spectroscopy or X-ray information. The resulting samples will be of crucial importance in the study of structure formation models, allowing to scan the structure history from early stages to their present state.

\begin{acknowledgements}
We would like to thank A. Iovino for making available an implementation of the EISily algorithm. We also thank the anonymous referee and all the VIRMOS team for useful comments.
\end{acknowledgements}

\end{document}